# Electronic phase separation in insulating (Ga,Mn)As with low compensation: Super-paramagnetism and hopping conduction

Ye Yuan[1, 2, *], Mao Wang[1, 2], Chi Xu[1,2], René Hübner[1], Roman Böttger[1], Rafal Jakiela[3], Manfred Helm[1,2], Maciej Sawicki[3], and Shengqiang Zhou[1] (s.zhou@hzdr.de)

[1] Helmholtz-Zentrum Dresden Rossendorf, Institute of Ion Beam Physics and Materials Research, Bautzner Landstrasse 400, D-01328 Dresden, Germany

[2] Technische Universität Dresden, D-01062 Dresden, Germany

[3] Institute of Physics, Polish Academy of Sciences, Aleja Lotnikow 32/46, PL–02668 Warsaw, Poland

Abstract:

In the present work, low compensated insulating (Ga,Mn)As with 0.7% Mn is obtained by ion implantation combined with pulsed laser melting. The sample shows variable-range hopping transport behavior with a Coulomb gap in the vicinity of the Fermi energy, and the activation energy is reduced by an external magnetic field. A blocking super-paramagnetism is observed rather than ferromagnetism. Below the blocking temperature, the sample exhibits a colossal negative magnetoresistance. Our studies confirm that the disorder-induced electronic phase separation occurs in (Ga,Mn)As samples with a Mn concentration in the insulator-metal transition regime, and it can account for the observed superparamagnetism and the colossal magnetoresistance.

* y.yuan@hzdr.de

Dilute ferromagnetic semiconductors (DFSs) have been of great interest due to their potential for spintronic devices [1-3]. However, even for the most studied DFS (Ga,Mn)As, understanding of the interplay between localization and magnetism in the insulator-metal transition regime still remains in a nascent stage. According to the Zener model, itinerant valence band holes mediate ferromagnetism between local Mn moments through *p-d* coupling [2, 4], establishing a long-range ferromagnetism in metallic (Ga,Mn)As. However, the observed ferromagnetism in (Ga,Mn)As samples exhibiting hopping conductivity acts contradictory to the above-mentioned itinerant-carrier mediated ferromagnetism mean-field model [5-7]. As a consequence, an impurity band model was developed and the ferromagnetism was explained by the double exchange interaction mechanism [8]. Thus, understanding the mechanism of hole-mediated ferromagnetism in (Ga,Mn)As, particularly on the insulating side of the insulator-metal transition, will improve our understanding of the carrier-mediated ferromagnetism in DFSs.

Different from the metallic (Ga,Mn)As samples which exhibit global ferromagnetic behavior, for the (Ga,Mn)As samples with Mn concentrations (around 1%) on the insulator regime of insulator-metal transition [9], an electronic phase separation was experimentally observed by low-temperature scanning tunneling microscopy (STM) [10] and theoretically explained by nuclear quantum effects [4]. This electronic phase separation causes a phase mixture of nano-sized ferromagnetic volumes (hole-rich regions) bubbling up in the otherwise paramagnetic matrix (hole-depleted regions) [11, 12].

In the present work, we report a systematic magnetic and electrical studies of insulating (Ga,Mn)As with the Mn concentration below the critical value of the insulator-metal transition. Our results confirm the picture that in insulating (Ga,Mn)As due to the electronic phase separation the holes mediated ferromagnetism is effective only on mesoscopically small distances what leads to a (blocked) superparamagnetic properties of the material.

(Ga,Mn)As samples for this study were prepared by implanting Mn ions into semi-insulating GaAs, followed by pulsed laser melting (PLM). The implantation was performed at the Ion Beam Center of Helmholtz-Zentrum Dresden-Rossendorf. The implantation energy was set to 100 keV, and the wafer normal was tilted by 7 degree with respect to the ion beam to avoid channeling. The Mn implantation fluences were $2\times10^{15}$ and $6\times10^{15}$ cm$^{-2}$ in two (Ga,Mn)As samples, denotes as G1 and G2, respectively. According to the stopping and range of ions in matter (SRIM) simulation[13], the longitudinal straggling ($\Delta R_P$) for the Mn distribution in GaAs is 31 nm. As a consequence, the thickness of Mn doped layers in GaAs as amounts to $2\Delta R_P = 62$ nm. A Coherent XeCl laser (with 308 nm wavelength and 28 ns pulse duration) was employed to recrystallize samples, and the energy density was optimized as 0.30 J/cm$^2$ to achieve the highest crystalline quality. The laser's pulse only melts several hundred nanometers below the surface, while the rest of the substrate stays at nearly ambient temperature. In this case, the huge temperature gradient between the molten thin layer and the substrate renders ultrafast recrystallization with a speed of several meters per second, which is much faster than the Mn diffusion speed in the layer. Such a process effectively denies Mn segregation or agglomeration. As proven recently [14-16] a careful optimization of the PLM condition is a prerequisite for obtaining high-quality DFS layers by the ion implantation method. Two Mn concentrations of 0.7% and 1.4% were determined by the secondary ion mass spectrometry (SIMS) using a Cameca IMS 6F micro-analyser. Magnetic properties were measured in a Quantum Design MPMS XL Superconducting Quantum Interference Device (SQUID) magnetometer equipped with a low field option. All magnetic data presented here have their relevant diamagnetic contributions evaluated at room temperature and subtracted accordingly[17]. It is worth noting that in some cases, the diamagnetic signal from the substrate largely influences the results when the ferromagnetic signal is weak. A GaAs holder which was intentionally mounted during the measurement effectively solved this problem by subtracting the background signal automatically during the measurement[18]. Herein, the GaAs strip holder was not used. Despite the fact that the absolute value of diamagnetic moment ($-2.5 \times 10^{-5}$ emu) from the GaAs substrate is around five times larger than the one ($4.0 \times 10^{-6}$ emu) of (Ga,Mn)As layer at around 10 K, the difference of GaAs substrate diamagnetic susceptibility between 10 K and 300 K is only 1.2%[19]. This indicates that the changing of diamagnetic signal with temperature contributes negligibly to the background subtraction. The detailed subtracting procedure is given in the Supplementary information. For all magnetic measurements, the magnetic field was applied along the in-plane [1$\bar{1}$0] direction. The normalized magnetization was calculated through dividing the measured magnetic moment from SQUID measurement by the Mn doped GaAs layer's volume which was calculated through multiplying the sample area by the layer thickness. For the temperature-dependent magnetization measurement, the continuous mode was used and the temperature changing rate was 3 K/min. The temperature and field dependent transport measurements were carried out by using van der Pauw geometry in a Lake Shore Hall Measurement System. Before transport measurements, samples were dipped in HCl solution to remove the surface oxide layer.

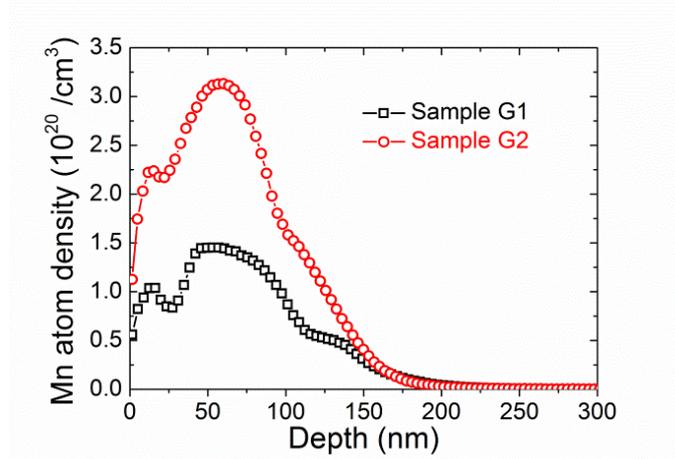

Fig. 1 Mn depth profiles in samples G1 and G2 determined by secondary ion mass spectrometry.

Atomic concentrations of manganese in sample G1 and G2 were determined by secondary ion mass spectrometry. As shown in Fig. 1, the Mn distribution is approximately Gaussian in both samples, and the peak values of Mn atom densities are around $1.5 \times 10^{20}$ and $3.1 \times 10^{20}$ cm$^{-3}$ for samples G1 and G2, respectively. Since there are $2.2 \times 10^{22}$ cm$^{-3}$ cations in zinc-blende GaAs, these numbers correspond to around 0.7% and 1.4% in these samples.

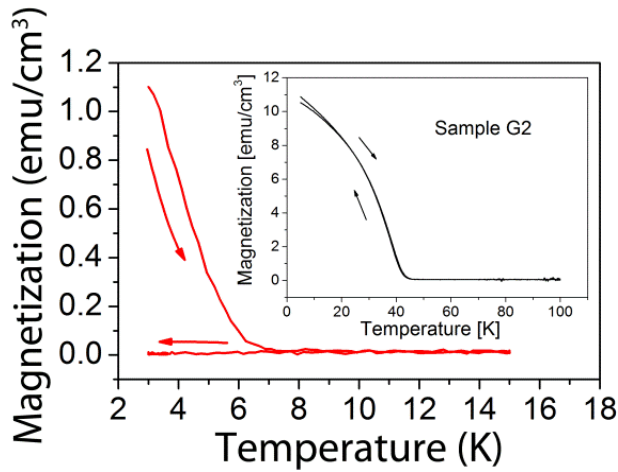

Fig. 2 (Color online) Temperature dependent thermo-remnant magnetization after the sample was cooled down to 3 K under a field of 1 kOe, and a following cool down measurement under zero field for sample G1. The inset shows the results using the same measuring procedure for sample G2. [9]

The basic information about magnetic constitution of the samples is inferred from low field, shown in Fig. 2. For thermo-remnant magnetization (TRM) measurement, the sample was cooled down under a field of 1 kOe, then at the base temperature the field was switched off by using a soft quench of the SQUID's superconducting magnet and the system was warmed up while collecting data. We note that for sample G1 the TRM exhibits a different behavior from a concave one which is characteristic for mean-field behavior exhibited in sample G2. Normally, for as-grown samples prepared by low temperature molecular beam epitaxy (LT-MBE), Mn interstitials and As antisites are present and act as double donors which compensate holes [20, 21].

Therefore, for LT-MBE grown samples with low Mn concentrations, a $T_C$ below 10 K is rare due to defect compensation. After the temperature reached 15 K which is above the magnetic transition temperature of 7 K, the sample G1 was re-cooled to the starting temperature at the same zero-field condition while the data recording was kept. Without applying any field, a zero-moment as shown in Fig. 2 excludes a global ferromagnetism in this sample. Differently, in the sample G2 which is doped with 1.4% Mn (the inset to Fig. 2), the overlapping of warming and cooling TRM curves indicates a spontaneous global ferromagnetic coupling throughout the whole layer. The distinct magnetic behavior under zero field conditions reveals a different range of magnetic coupling in (Ga,Mn)As doped with Mn concentrations on both sides of the Anderson-Mott insulator-metal transition.

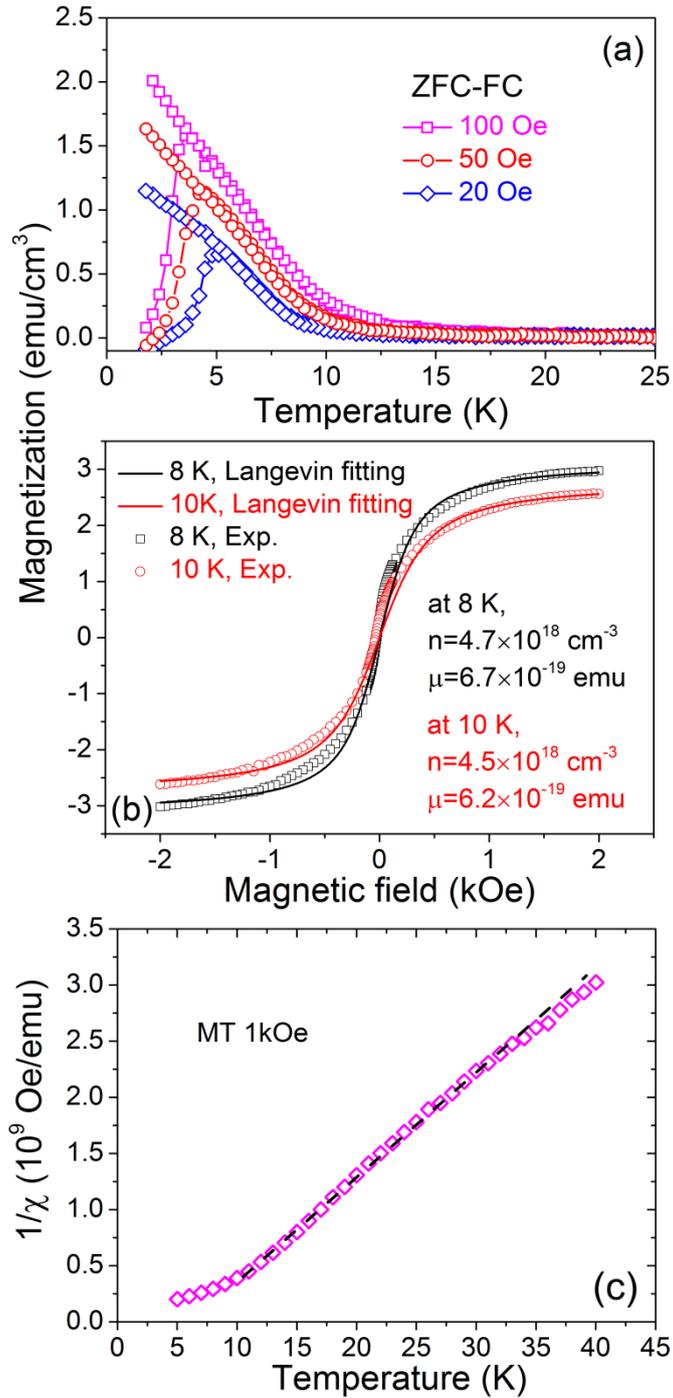

Fig. 3 (Color online) (a) Field cooling and zero field cooling temperature dependent magnetization curves measured under fields of 20, 50, and 100 Oe in sample G1. (b) Experimental (open squares) and Langevin fitting (solid lines) results of magnetization vs. magnetic field curves at 8 and 10 K of sample G1. (c) Inverse magnetic susceptibility (diamonds) versus temperature of sample G1 under a field of 1 kOe. The dash line shows a linear dependence above the $T_B$.

To obtain more details about magnetic properties in sample G1, we have performed low-temperature sample cycling using weak fields in the well-established protocol of zero-field cooling (ZFC) and the field cooling (FC). Accordingly, a typical superparamagnetic (SPM) behavior is observed, indicating that the magnetic signal comes from the nano-sized ferromagnetic clusters in sample G1. As shown in Fig. 3(a), a bifurcation is present between ZFC and FC curves where a maximum (the mean blocking temperature $T_B$) shows up in the ZFC curve. Moreover, $T_B$ shifts to lower temperature upon increasing applied fields. The above-mentioned phenomena reveal a nature of blocking superparamagnetic behavior at low temperature and low magnetic fields: after cooling down under zero-field condition, moments of nano-sized magnetic grains are randomly frozen. However, the applied field overcomes part of the blocking energy barrier and immediately polarizes a fraction of moments. Upon increasing temperature, more magnetic moments are aligned due to the thermo-agitation, resulting in the detected increased magnetization. On the other hand, a larger applied field would polarize more moments with higher blocking energy barrier, leading to the low-temperature-side shift of $T_B$ [22-24].

Magnetic properties of the superparamagnetic system could be well described by a Langevin function in Eq. (1) [25].

$$M(T) = n\mu[\coth\left(\frac{\mu_0 \mu H}{k_B T}\right) - \frac{k_B T}{\mu_0 \mu H}] \qquad (1)$$

where $n$ is the number of single domain particles per unit volume, $\mu$ is the magnetic moment of a single magnetic particle, $\mu_0$ is the magnetic permeability of vacuum, and $k_B$ is the Boltzmann constant. As presented in Fig. 3(b), fittings by the Langevin function reproduce the magnetic field dependent magnetization experimental data at 8 and 10 K, respectively. However, the deviation between fitting curves and experimental results indicates a weak dipolar interaction in the system, which was shown by the inverse magnetic susceptibility versus temperature curve in Fig. 3(c). The magnetic cluster density $n$ and magnetic moments per magnetic particle $\mu$ at 8 and 10 K are fitted as $4.7 \times 10^{18}$ cm$^{-3}$ and $6.7 \times 10^{-19}$ emu (~72 $\mu_B$), and $4.5 \times 10^{18}$ cm$^{-3}$ and $6.2 \times 10^{-19}$ emu (~67 $\mu_B$), respectively. Considering that each ferromagnetic Mn atom contributes 4 $\mu_B$ magnetic moments [20], each ferromagnetic cluster consists of around 18 and 17 Mn atoms at 8 and 10 K, respectively. By considering the atom density of GaAs which is $4.2 \times 10^{22}$ /cm$^3$, the average Mn atom concentration which is enclosed in the ferromagnetic nano-clusters is 0.4%. In addition, we employ the standard formula for the dynamical blocking, $KV = 25k_B T$, where $K$ is the anisotropy constant in (Ga,Mn)As (5000~50000 erg/cm$^3$)[26], $k_B$ is the Boltzmann constant, $V$ is the volume of the magnetic cluster, and the 25 factor is decided by the experimental time scale which is around 100 seconds in the SQUID magnetometry. A diameter of one magnetic cluster is calculated between 10 and 20 nm at 5 K, what is in line with the low-temperature STM studies, by which one can infer hole-rich clusters of 4.6 to 7.2 nm[10]. Interestingly they are lower than those elaborated recently from magnetic studies performed on low Mn concentration doped, below Mott-critical hole concertation, core-shell (Ga,Mn)As nanowires exhibiting similar superparamagnetic-like properties[27, 28]. The inverse magnetic susceptibility versus temperature curve of sample G1 [shown in Fig. 3 (c)] approximately presents a linear dependence above $T_B$, however the extrapolated Curie-Weiss fit suggests a weak interaction between ferromagnetic nanoclusters.

Actually, superparamagnetic properties of (Ga,Mn)As were seen before. However, the observed SPM properties were normally caused by the generated MnAs second phase resulting from the spinodal-decomposition[29-31]. Differently, no signature of Mn segregation appears in sample G1[9], thus the nano-sized ferromagnetic clusters are expected to result from the fluctuation of local hole density, namely the electronic phase separation taking place at the localization boundary [10]. To substantiate the electronic phase separation concept, the saturation magnetization of sample G1 is measured at 2 kOe at 5 K. In this case, the applied 2 kOe field is sufficient to totally polarize the ferromagnetic coupled volumes (as indicated in Fig. 3b), with the remaining paramagnetic Mn atoms can contribute only a moment of 0.63 $\mu_B$ according to the Brillouin function calculation. Through integrating SIMS results along the depth profile, one can obtain the Mn atom numbers in the sample. As a result, the magnetic moment per Mn atom was calculated through dividing the magnetization by the integrated Mn atom numbers. The magnetic moment per Mn atom is normalized to be 2.3 $\mu_B$, which is just between the 3.5~4 of ferromagnetic Mn atoms and 0.63 $\mu_B$ of paramagnetic ones[20]. This indicates that only part of the Mn atoms are involved in the ferromagnetic phase, and it also can be understood by considering the co-existence of the nano-sized hole-rich ferromagnetic bubbles and the hole-deplete paramagnetic matrix. In addition, high-resolution transmission electron microscopy (TEM) imaging of sample G2 together with spectrum imaging based on energy-dispersive X-ray spectroscopy in the scanning TEM mode exhibits neither Mn-rich nor MnAs clusters, as shown in Figure S2 in the supplementary information. It indirectly excludes the possibility that Mn-rich clusters cause the SPM in sample G1 since Mn-clustering should first occurs in samples with higher Mn concentration. It is worth noting that the inhomogeneous distribution of Mn along the depth as shown in Figure 1 and Figure S2. The deeper region with much lower Mn concentration could contribute some paramagnetic signal. In addition to the spin-disorder and thermo-fluctuation contribution, nuclear quantum effects are also possible to result in such an electronic phase separation in the sample which is in the critical metal-insulator transition regime [4].

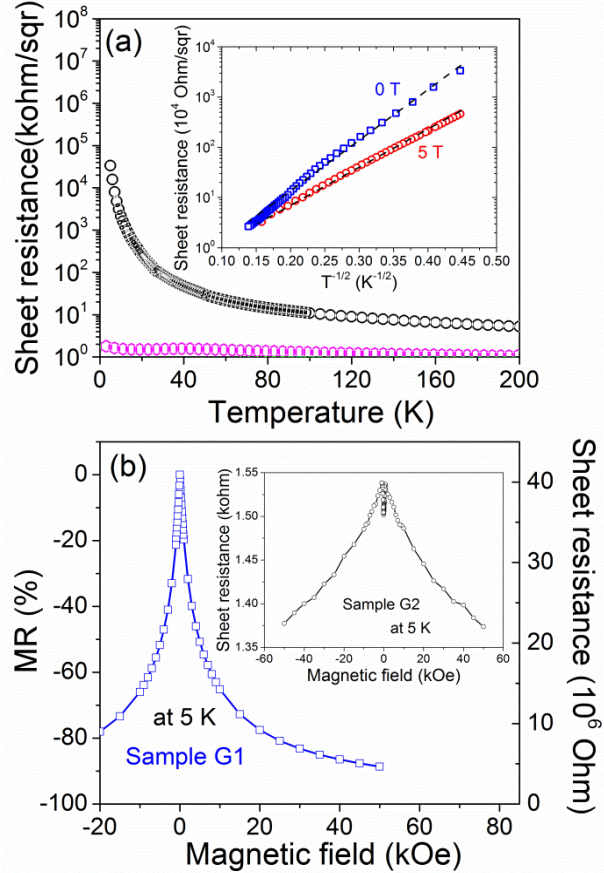

Fig. 4 (Color online) (a) Temperature dependent resistance under zero field of sample G1 (circles) and G2 (hexagons). The insert shows the $T^{-1/2}$ dependent resistivity: Experimental results (open squares and circles) and the fitting curves by Eq. (2) (solid line). (b) Experimental results (open squares) of field dependent negative magnetoresistance of sample G1 at 5 K. The inset shows negative magnetoresistance between ±5 T at 5 K of sample G2.

According to the Zener model, itinerant valence band holes couple Mn moments in metallic (Ga,Mn)As. However, in our (Ga,Mn)As sample G1, the resistivity increases upon decreasing temperature in all temperature range, presenting an insulating behavior, as shown in Fig. 4(a). The fitting of $T^{-1/2}$ dependent resistance from 5 to 40 K is given by Eq. (2):

$$\rho(T) = \rho_0 \exp\left(\frac{E}{k_B T}\right)^{\frac{1}{2}} \qquad (2)$$

where the pre-exponential constant $\rho_0$ and activation energy E are fitting parameters, and $k_B$ is Boltzmann constant. A linear $T^{-1/2}$ temperature dependence of resistivity indicates that the Efros-Shklovskii variable-range hopping (ES-VRH)[32] dominates the electrical transport, as shown in the inset to Fig. 4(a). It is seen there that the field of 5 T reduces the slope, indicative that the hopping energy barrier is reduced by the magnetic field from 2.0 to 1.4 meV. When such hopping barrier decreases, the ES-VRH can occur much easier, giving rise to a colossal magnetoresistance (MR), as high as 95% as observed at 5 K, presented in Fig. 4(b). Similar colossal negative MR was observed in oxide materials [33], and an electronic separation model has been developed by considering the competition between carrier-mediated ferromagnetism and intrinsic anti-ferromagnetism [11]. However, in the metallic (Ga,Mn)As sample G2, the negative magnetoresistance is only 11%, resulting from the quantum localization effect [34] which also contributes, even not as much as the ES-VRH effect, in sample G1. As shown in Fig. 4(a), sample G2 exhibits metallic behavior, the same as the high Mn concentration doped (Ga,Mn)As prepared by LT-MBE.

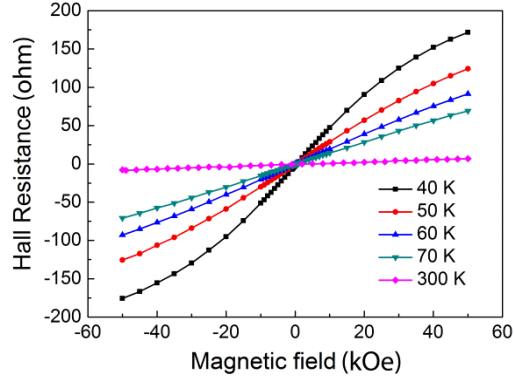

Fig. 5 (Color online) Hall resistance curves of sample G1 measured at 40, 50, 60, 70, and 300K.

The anomalous Hall effect (AHE) is observed in sample G1 at temperatures above $T_B$, confirming the coupling between spin-polarized carriers and local magnetic moments. The positive slope of Hall curve at 300 K confirms the *p*-type conductivity induced by Mn doping. According to previous works [29, 35], in addition to global ferromagnetic samples, the AHE is also present in the system consisting of nano-sized ferromagnetic clusters.

In summary, an insulating (Ga,Mn)As with 0.7% Mn is prepared by ion implantation combined with pulsed laser melting. The hopping conduction dominates the electrical transport and the sample is on the insulating side of the insulator-metal transition. Disorder induced electronic phase separation causes superparamagnetic behavior with an 8 K blocking temperature, which is consistent with the calculation by Bae et al. [4].

Support by the Ion Beam Center (IBC) at HZDR is gratefully acknowledged. This work is funded by the Helmholtz-Gemeinschaft Deutscher Forschungszentren (HGF-VH-NG-713). The author Y. Y. thanks financial support by Chinese Scholarship Council (File No. 201306120027). Financial support by the EU 7th Framework Programme under the project REGPOT-CT-2013-316014 (EAgLE) and the international project co-financed by Polish Ministry of Science and Higher Education, Grant Agreement 2819/7.PR/2013/2, is also gratefully acknowledged. Furthermore, the funding of TEM Talos by the German Federal Ministry of Education of Research (BMBF), Grant No. 03SF0451 in the framework of HEMCP is acknowledged.

# Supplementary information for

"Electronic phase separation in insulating (Ga,Mn)As with low compensation: Super-paramagnetism and hopping conduction"

Ye Yuan[1, 2, *], Mao Wang[1, 2], Chi Xu[1,2], René Hübner[1], Roman Böttger[1], Rafal Jakiela[3], Manfred Helm[1,2], Maciej Sawicki[3], and Shengqiang Zhou[1]

[1] Helmholtz-Zentrum Dresden Rossendorf, Institute of Ion Beam Physics and Materials Research,
Bautzner Landstrasse 400, D-01328 Dresden, Germany
[2] Technische Universität Dresden, D-01062 Dresden, Germany
[3] Institute of Physics, Polish Academy of Sciences, Aleja Lotnikow 32/46, PL–02668 Warsaw, Poland

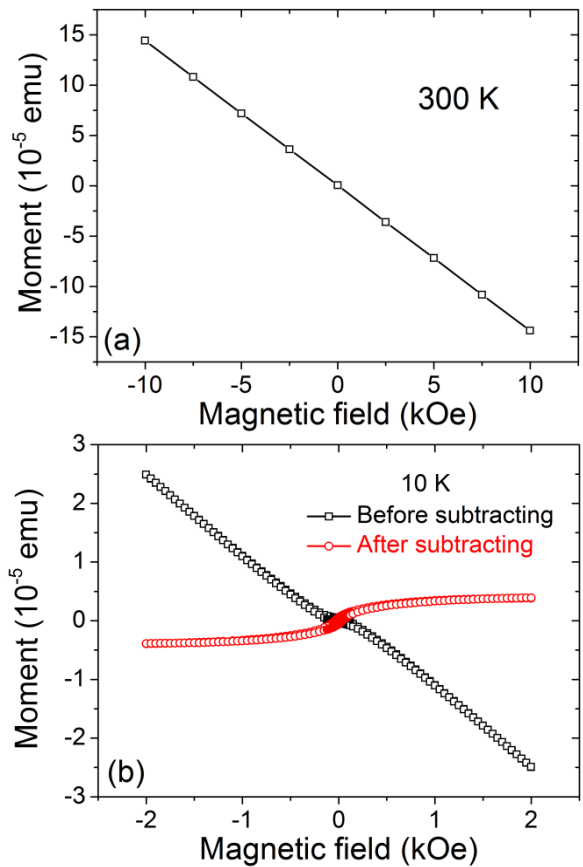

Figure S1. The magnetic field dependent magnetic moment for sample G1 at (a) 300 K and (b) 10 K.

For the diamagnetic signal subtracting, the MH curve of sample G1 was firstly measured at 300 K to obtain the diamagnetic susceptibility ($1.44\times10^{-8}$ emu/Oe) of GaAs substrate. Then the obtained susceptibility was used to subtract the diamagnetic background signal at 10 K. As shown in Figure S1(b), after subtracting the diamagnetic signal, the susceptibility of the MH curve becomes positive.

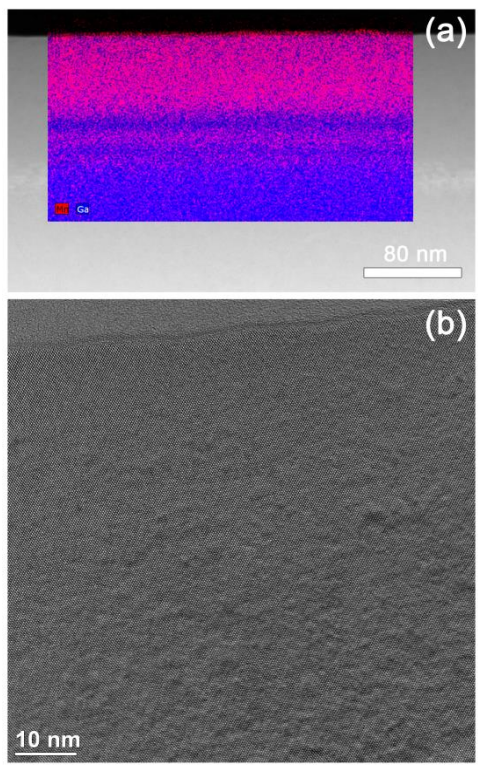

Figure S2. Transmission electron microscopy (TEM) images of sample G2 (a) High-angle annular dark-field scanning TEM image together with the Mn (red) and Ga (blue) element distributions obtained by energy-dispersive X-ray spectroscopy (EDXS). (b) High-resolution TEM micrograph of the surface region of sample G2.

In Figure S2, we show TEM analysis results of the 1.4%-Mn-doped sample G2 which should have a higher potential to present manganese clusters than the 0.7%-Mn-doped sample. However, neither Mn-rich clusters nor MnAs clusters are observed in the high-resolution TEM micrographs or the element distributions obtained by spectrum imaging based on EDXS in scanning TEM mode. The Mn distribution is laterally uniform. The inhomogeneity along the depth, particularly in the deeper region, is a result of the ion implantation profile and the redistribution during the regrowth from melting.